# Digital health shopping assistant with React Native: a simple technological solution to a complex health problem

Alina Govoruhina & Anastasija Nikiforova

**Abstract**
Today, more and more people are reporting allergies, which can range from simple reactions close to discomfort to anaphylactic shocks. Other people may not be allergic but avoid certain foods for personal reasons. Daily food shopping of these people is hampered by the fact that unwanted ingredients can be hidden in any food, and it is difficult to find them all. The paper presents a digital health shopping assistant called "Diet Helper", aimed to make life easier for such people by making it easy to determine whether a product is suitable for consumption, according to the specific dietary requirements of both types - existing diet and self-defined. This is achieved by capturing ingredient label, received by the app as an input, which the app analyses, converting the captured label to text, and filters out unwanted ingredients that according to the user should be avoided as either allergens or products to which the consumer is intolerant etc, helping the user decide if the product is suitable for consumption. This should make daily grocery shopping easier by providing the user with more accurate and simplified product selection in seconds, reducing the total time spent in the grocery stores, which is especially relevant in light of COVID-19, although it was and will remain out of it due to the busy schedules and active rhythm of life of modern society. The app is developed using the React Native framework and Google Firebase platform, which makes it easy to develop, use and extend such solutions thereby encouraging to start actively developing solutions that could improve wellbeing.

## Introduction

Nowadays, more and more people are reported as having an allergy, which can range from very simple reactions closer to discomfort to anaphylactic shock. Allergies affect both social, emotional, and financial wellbeing [1]. Back in 2016, the European Academy of Allergy and Clinical Immunology (EAACI) confirmed that allergy is the most common chronic disease in Europe, where up to 20% of allergy patients struggle daily with fear of a possible asthma attack, anaphylactic shock, or even death from an allergic reaction [2]. Moreover, more than 150 million Europeans suffer from chronic allergic diseases, and it is predicted that by 2025 half of the entire EU population will be affected by them.

Recent studies report that prevalence surveys, healthcare utilization data, and results of longitudinal cohort studies around the world indicate that food allergy is a growing societal burden [3]. They also found that in the USA, a population-based cross-sectional prevalence survey of more than 50 000 households estimated that food allergy affects about 1 in 10 adults [4] and 1 in 12 children [5]. They indicate that more than 10% of the US population suffers from at least one food allergy, with even more people reporting a current food allergy in the absence of conclusive symptoms. For example, in this survey, nearly 20% adults reported that they had at least one current food allergy, and more than 11% of children had a parent-reported allergy. The eight foods that cause the most food allergy reaction are milk, soy, eggs, wheat, peanuts, tree nuts, fish and shellfish. Although food is only one of possible sets of allergens, Wood et al. [6] found food to be one of the most common triggers for anaphylaxis, a life-threatening reaction, along with medicines / drugs and insect stings/ bites.

Food allergies continue to affect a large proportion of children and adults worldwide, causing the economic impact of food allergies on individuals and society that has become a priority for many stakeholders, including individuals, families, schools, workplaces, healthcare systems, the food industry, and policymakers [1, 7-8]. While the health risks are probably the most important, allergies also have economic consequences. Economic studies evaluating the costs of food allergy seek to ultimately quantify direct costs such as medications and indirect costs such as restricting employment (e.g., to care for a child with food-allergies), and intangible costs such as quality of life [1, 9]. Gupta et al. [10] estimated that childhood food allergies cost approximately $25 billion per year in the US. Living with a food allergy is associated with an annual economic cost more than $4000 per child, in addition to risks of anxiety and depressive symptoms, negatively impacting the quality of life and mental health [11].

Allergies can be dealt with in two ways - prevention and treatment, where prevention is always a priority. This leads to the development of appropriate diets. Other people, however, may not be allergic but avoid certain foods for personal reasons, such as a sport diet etc. All in all, many consumers avoid certain products or ingredients. Therefore, supportive solutions to facilitate shopping for such people and providing recommendations on the suitability of the food for their lifestyle would be beneficial both in view of the active rhythm of humanity, which makes it difficult to view the entire offer of products, and their manual filtering, which can lead to inaccurate results, when a particular restricted or avoidable ingredient was not noticed. This is even more the case in times of COVID-19 pandemic when such shopping activities are likely to be reduced in time as much as possible.

We suppose that ICT, the trends of digitalization and some aspects of Industry 4.0 can play the role of a preventor, without requiring a lot of resources due to the advances in the software development domain. The solution proposed in this paper is a very simple mobile application, which falls into this category - it provides the user with recommendations regarding consumed product ensuring the opportunity to prevent allergic reactions by excluding the allergens consumed. This m-health application is intended for users who refuse to use any ingredient in the diet, for health, ethical or other reasons. This simple app provides the user with a list of seven predefined diets: vegan, vegetarian, pescovete, or diets that prohibit the use of sugar, gluten, dairy, or nuts. This list of prohibited dietary ingredients, is obtained from online health magazines. People following one or more of these diets are considered the main audience of the app, although in addition to these dietary options, the user can add additional personal ingredients to their profile that should be excluded from their diet, i.e., being informed about their presence in food. Daily grocery shopping by these people is hampered by the fact that unwanted ingredients can be hidden in any food, and it is difficult for a person to remember and find them all. The app is currently intended for Anrdoid smart device users, requiring an internet connection and a camera.

While currently, the app is specifically designed to scan and process a list of food ingredients by searching for food ingredients, it can be adapted to the composition of e.g., cosmetic products in the future by extending its source code or developing alternative solutions following a simple procedure presented in this paper, constituting a call for action. It can also be used to track the safety of the product, indicating the list of ingredients or foods hazardous to humans, thereby contributing to the food safety being a part of Sustainable Development Goals (SDG).

The paper is structured as follows: Section 2 refers to the background and motivation of the study, Section 3 presents the developed application, while Section 4 presents currently limitations and future works implying from them, and Section 5 concludes the paper.

## Background and Motivation

The fast pace of life does not allow to spend much time manually analysing the list of ingredients in grocery stores. This process could be improved by offering an application that can determine the composition of the product, returning to the user the result of the suitability of a given product for diet and preferences of a particular user thereby substituting recommender / mobile shopping assistant.

This paper presents such an application titled Diet Helper. The app recognizes ingredients in the ingredient list photo that are not recommended for use according to user-selected diets and added unwanted ingredients. However, it is not the first solution in this domain. Therefore, it is important to understand what makes it different from already existing. Most similar apps use barcodes to select foods - [12], *HealthMe*, *ContentChecked*, *MyFoodFacts* etc. Another cluster of solutions allows the person to understand, which cuisine to choose - *iEatOut Gluten Free and Allergen Free* – recommending on which restaurant is the most allergic-tolerant by considering so-called allergy-friendly rating – *AllergyEatsMobile*. These apps, however, are limited to the region for which the app is intended. This means, that in most cases these apps require maintaining data, e.g., for all possible products and their barcodes or connecting to the source and searching for a particular artifact (e.g., database or scrapping the entire internet). One more cluster of solutions is closely related and even based on the citizen science, including but not limited to crowdsourcing. This also often include a reference to or is used in combination with the open data. Several examples, which are gaining an increased popularity are Spanish solution called App Plantes - "*Flora urbana y alergia, ¿cooperas?*" (*Urban flora and allergy, do you cooperate?*) aimed at informing citizens about allergenic plants in the region and the level of the allergy risk they pose, where the latter depends heavily on their phenological state, i.e., closed flower, open flower and/or fruit. It is expected that the users - citizens collaborate in the collection of these data contributing to the development of a map on the state of plants, i.e., a citizen answers a short questionnaire and submit a photo of each plant. These data are aggregated and made openly available via an interactive map showing the different levels of allergen alerts. While *Miils* is a Finish open data-based application that finds recipes with nutrition and allergy information, which is the used to create own recipes based on the products the person can have[1]. Considering the scope and nature of these solutions they both require maintenance of the database with many of them requiring either constant or on-demand scrapping of the Internet thereby keeping the database up-to-date, which, however, negatively affects the performance of these solutions as well as their effect on the battery of the smart device in use. Unlike these solutions, the simplicity of *Diet Helper* does not require maintenance of data for all possible products and their barcodes, which will also depend on the desire of user to supplement these data with the correct food information. Instead, *Diet Helper* analyses the list of captured / photographed food ingredients, comparing each ingredient to the prohibited dietary ingredients in the database of limited size to determine if the product is avoidable (or desirable). This makes the processing of each request and respective recommendation made by the app sufficiently faster (the handling of unwanted ingredients will be described below). In addition, the results are expected to be more accurate, since a label typically contains more authoritative information provided by the actual producer, where the risk of the ingredient to be missed is lower.

## Diet Helper

The functionality of the app is intended for authorized users. In other words, the only feature available to unregistered users is to register in the system - the app provides authentication with email as well as *Facebook* and *Google* authentication supported by *Firebase*.

Once the user is registered and logged in the system, the user can (1) select one or more diets from the proposed list: (1.1) vegan, (1.2) vegetarian, (1.3) pesco vegetarian, (1.4) gluten-free, diets (1.5) sugar-free, (1.6) milk-free, (1.7) nut-free, or (2) add self-defined ingredients to the profile that are not described in the above listed diets automatically forming a "Custom" diet. In addition, it is possible (3) to view the description of the diets used in the app and the list of forbidden ingredients in the diet, which is complemented by other management features such as changing the language of the application interface, changing the username etc.

---

[1] https://datos.gob.es/en/blog/citizen-science-and-open-data-help-society

Then, in accordance with the main functionality and purpose of the app, the user is expected to take a photo of the ingredient label – a list of food ingredients, or upload it from local storage, receiving back the text content of the processed photo in text format as the result of photo processing by the app. This is done by an external *React Native Text Detector* module that converts an ingredient label / image into text without additional programming efforts at the developer end.

Then, if successful, the user receives confirmation as to whether the processed list of ingredients complies with the diet(s) selected by the user. In case there is an ingredient, which is listed in one or more lists (selected or defined diets), the user receives a warning along with a list of diets whose forbidden ingredients were found. The list of predefined dietary ingredients, list of diets and their description is managed by the administrator, who can add, update, or delete them. The administrator is also allowed to manage the list of users, if necessary (e.g., by users' request). All in all, there are three user groups - unregistered user, registered user, and administrator (Fig. 1).

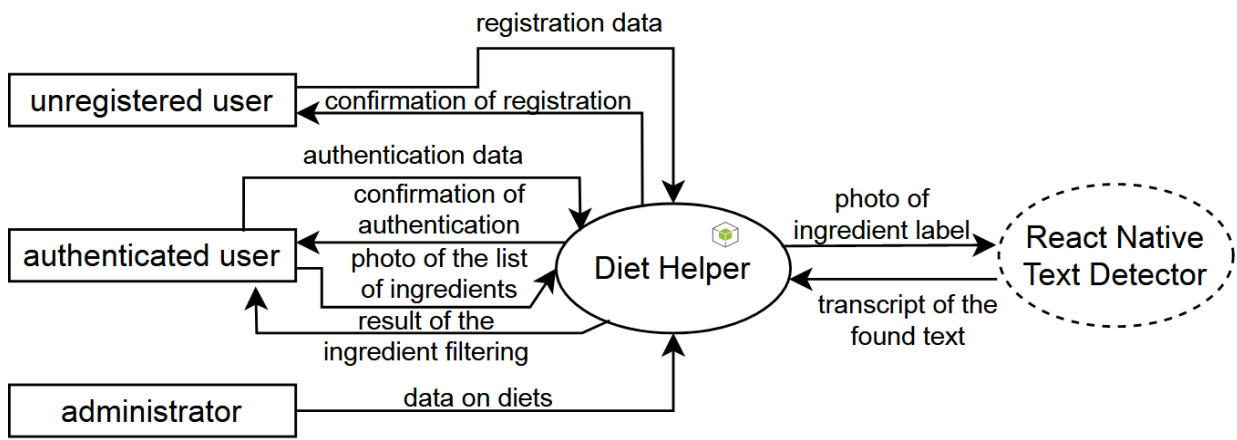

**Fig. 1.** Data flow chart, level 0

Let us briefly describe function design referring to the most important ones in the context of the app. Fig. 2 shows how the system works with user-selected diets to get the result of ingredient filtering. After registration, a new user is automatically registered in the system. If no diet is selected for the user, it can be selected, and if a diet or diets are added, the diet can be deleted from the list. The user can choose no diet and choose to take a picture or add a list of ingredients to a product. In this case, only the unwanted ingredients added by the user will be considered in the ingredient filtering. If the user has at least one, but not all, diets selected, another diet can be added to the list of selected diets. The user can choose not to add or delete the diet, but to take a photo or add an unwanted ingredient. The user can add its own ingredients to highlight them in the ingredient filtering process. If the user does not have such ingredients, only the prohibited ingredients of the selected diet(s) will be taken into account in the filtering. If the user has not selected any diet and has not added any unwanted ingredients, the filtering result will always show that no unwanted ingredients were found in the product.

If the user has chosen to take a list of ingredients and no textual information is recognized in the captured image, the user is asked to retake the photo. If the text is recognized, it is automatically filtered with the prohibited ingredients from the user's list of diets and the filter result is returned to the user. Similarly, if the text is not recognized on the uploaded image, the user can try to add the image again. The user chooses to take a list of ingredients with the camera of the smart device.

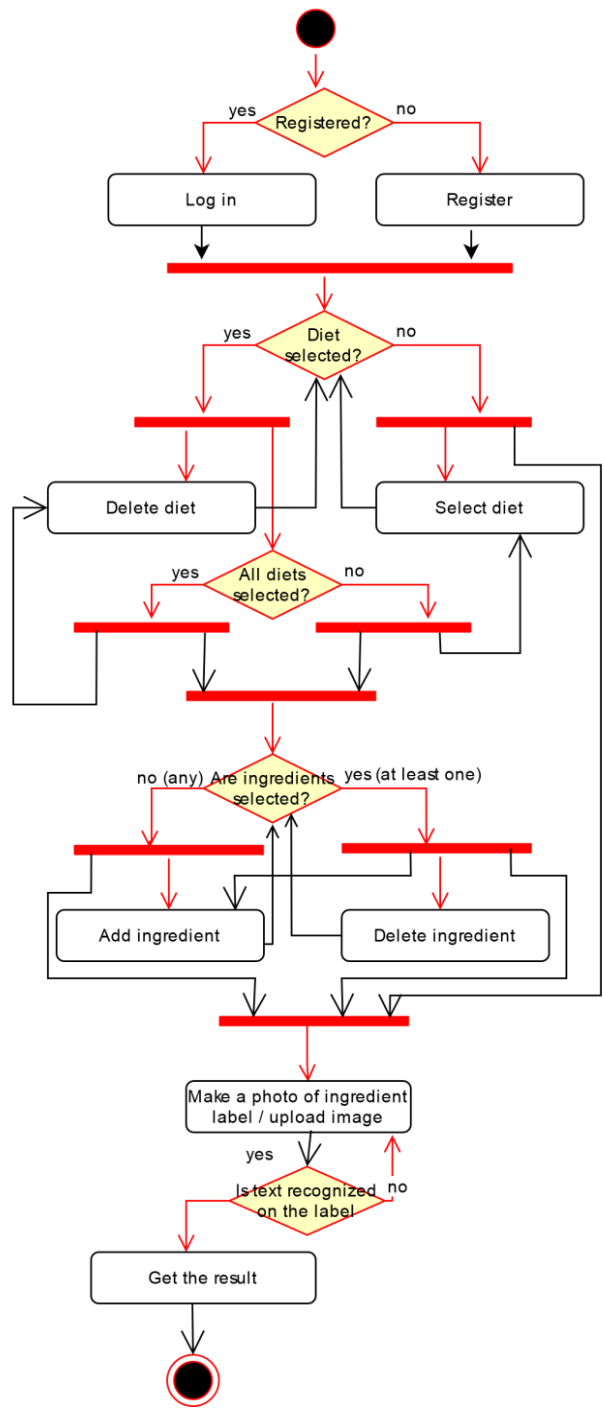

**Fig. 2.** Activity chart (ingredient filter module)

It is assumed that the app will use the shooting feature more often than downloading a picture. The photo is fed to the React Native Text Detector module, which retrieves the textual information from the image if it is found. The diagram shows the scenario in which text is recognized in an image. The transcript of the text together with the diets selected by the user and the unwanted ingredients entered are passed to the filtering module of the system. The result of the processing of the ingredients (highlighted unwanted ingredients found, if any) is returned to the user.

There are a few general requirements or prerequisites for the app to work correctly. First, the current version of the application requires an internet connection. This allows not to store the entire database on the users' smart device rather than redirecting to it upon request. Secondly, given the nature of the application, it is clear that the correctness of the result of the application depends on the recognition of the corresponding ingredient label and its text by *React Native Text Detector*. This works best if the captured ingredient list image has clearly visible text, which means that when the user decides to capture

/ take a photo of ingredient label, the camera must be of sufficient quality with recommended resolution above 150dpi, while the image should be well lit, and the text area should be in focus.

There are also some assumptions and dependencies. First, it is assumed that all banned ingredients in the diet are 100% forbidden. The system prevents the presence of ingredients of dubious origin, such as animal or plant / vegetable origin in the case of a vegan diet. The app ignores this and considers such ingredients to be non-organic. If an ingredient is part of the prohibited ingredients of one of the diets, then it certainly does not belong to that diet. It is, however, up to the user to decide whether to take into an account the app's warning, i.e., the app intends to help to make a data-driven decision.

## Diet Helper Database

The database of the app is divided into two parts, where the first part stands for diets and prohibited ingredients, and the other - for registered users. User data includes the selected name, unique email, encrypted password, user-selected diets, and personally added unwanted ingredients that the user wants to highlight in the product's ingredient list.

The database contains the following data on diets: the unique name of the diet, which is used as a key, a short description of the diet, and ingredients that are not allowed in the diet. Multiple prohibited ingredients may be assigned to the same diet, and the same ingredient may be prohibited in multiple diets. Registered users have the option to choose one or more diets, add one or more unwanted ingredients. As a result, these ingredients are searched in the food ingredient lists. The list of prohibited dietary ingredients, is obtained from online health magazines[2][3][4]. These sources were combined to enrich the result set ensuring its comprehensiveness and completeness. It forms the diet-related part of the database for the developed app.

Let us now elaborate on the database design. The *Firebase Cloud Firestore* NoSQL *d*atabase was selected for the app. *Firebase Cloud Firestore* provides fast data recording, i.e., 1-2 seconds, as well as offers to keep track of any data changes, so in the app it is also possible to immediately display the changes on the screen ensuring interactive operation of the system. The database is divided into two collections - *Users* and *Diets*.

User data are stored in two collections. The first stores authentication data - this is the Firebase authentication module. Authentication data includes e-mail, supported service: e-mail, Facebook or Google account and unique identifier - UID. The password is stored in an encrypted form with the *scrypt* hash. Remaining user information is stored in the *Users Firestore* collection.

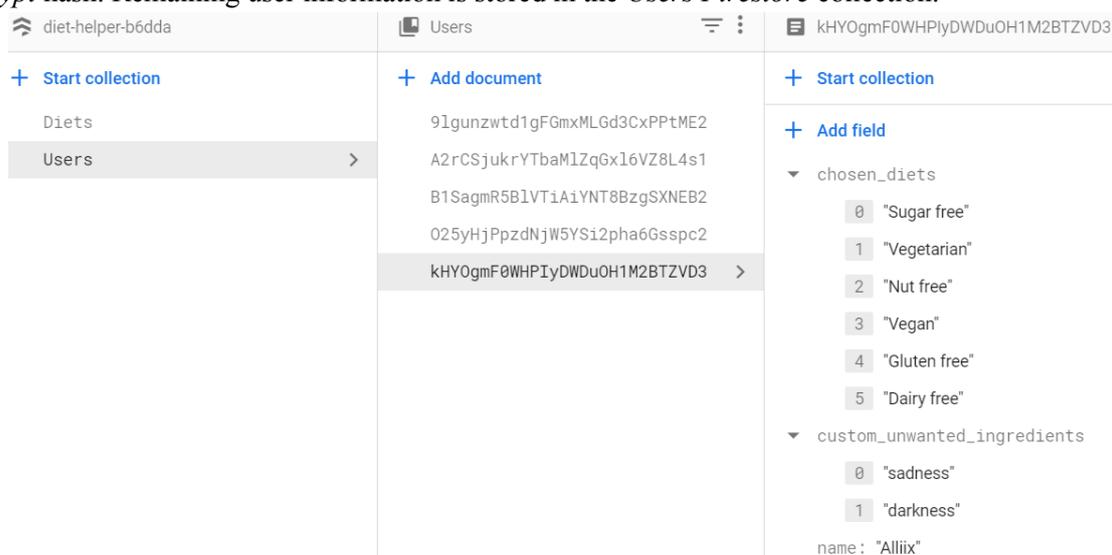

---

[2] https://www.webmd.com/diet/features/hidden-sources-of-gluten#1,
[3] https://www.virtahealth.com/blog/names-for-sugar,
[4] https://www.eatforhealth.gov.au/food-essentials/five-food-groups/lean-meat-and-poultry-fish-eggs-tofu-nuts-and-seeds-and

**Fig. 3.** Cloud Firestore Users collection data

The collection stores the user-specified name, the selected diet (chosen_diets), and the users' entered unwanted ingredients (custom_unwanted_ingredients). Each user's data are stored in a separate document, the key to which is the user's unique identifier UID (Fig. 3). This, i.e., storing each user's data in one document reduces the number of database requests, as all user data can be received in a single request for a single document.

Fig. 4 shows the data of the diet collection. The collection of diets contains documents about diets - each document is a diet, the unique identifier of which is the name of the diet.

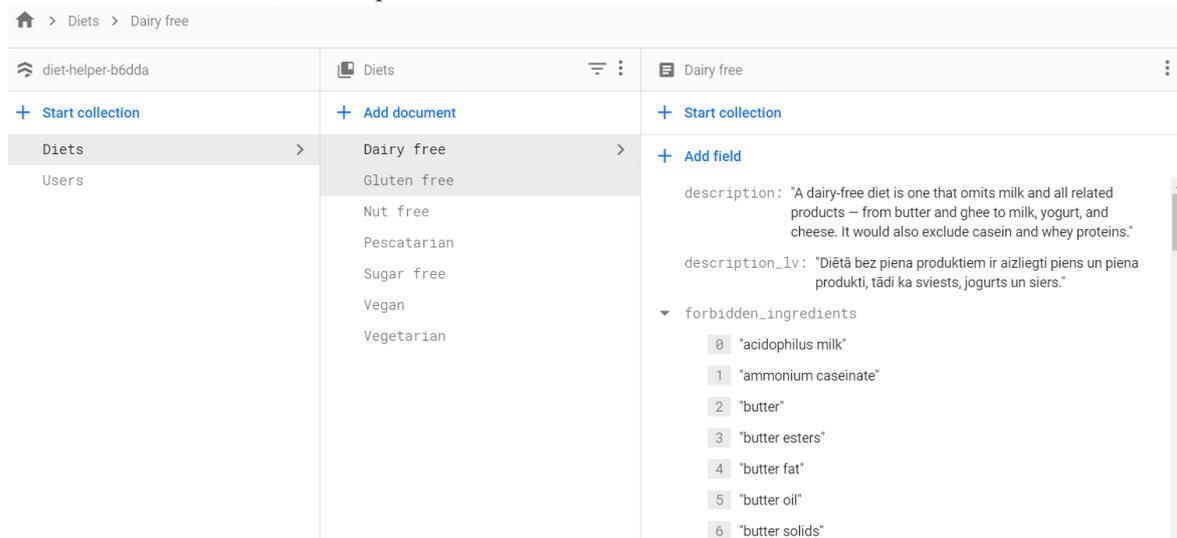

**Fig. 4.** Cloud Firestore Diets collection data

The diet document consists of three fields: a description (*description*) and the prohibited ingredients of the diet (*forbidden_ingredients*). The description fields store textual data - a general dietary description, while the prohibited ingredients field stores an array of textual information with all the ingredients that are not allowed in this diet. This type of storage was chosen to reduce the number of database requests. In other words, if the ingredient lists were also stored in collections, the request for each specified ingredient would be treated as a request for a separate document. By storing the prohibited ingredients in the text information array, the entire list of ingredients would be downloadable each time information about a diet is requested, as partial downloading of the *Cloud Firestore* document is not possible.

A request for an individual diet counts as a single request, with the app retrieving both the dietary descriptions and all of its prohibited ingredients. The app needs all the prohibited ingredients in the diet when processing the list of food ingredients, so it is better for the system to make a single request for all the dietary data.

### System modules

Similar to a database, the app is splitted in two modules: (1) a user module and (2) an ingredient list filtering module. The user module works with the system database and the external *Firebase* authentication module.

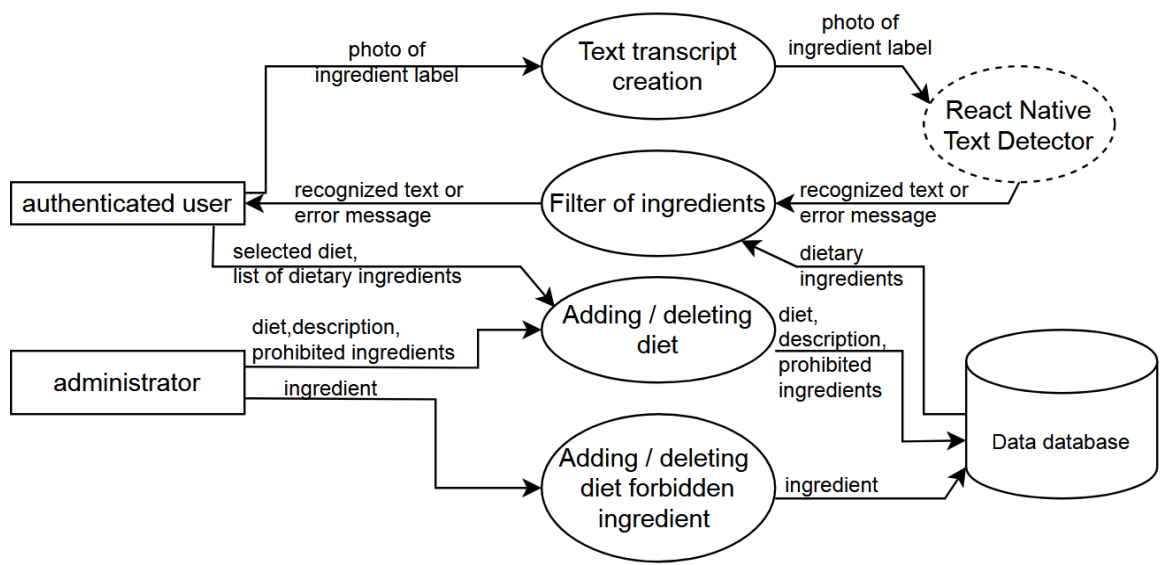

**Fig. 5.** Dataflow chart (level 2, ingredient filtering module)

User-selected diets and user-unwanted ingredients are fed to a text filtering module, determining which diets and ingredients to filter the list of ingredients in the user's image. In addition to communication with the text filtering module, the user module includes functions that determine the personalized operation of the application.

The ingredient list filtering module interacts with users, the database, and the external module - React Native Text Detector (Fig. 5). From the user module, the ingredient filtering module retrieves the diet(s) selected by the user, the unwanted/ prohibited defined by the user, and a photo with the list of ingredients of the product created or uploaded by the user. The text filtering module obtains data about the prohibited ingredients in the diet from the "Diet Database". The user-captured photo module passes the React Native Text Detector to an external system for conversion into text information that can be further processed by the filter module. The resulting text searches for ingredients that are prohibited in the diets and unwanted ingredients selected by the user. The image in the ingredient filter module is not processed. Instead, it is passed to an external React Native Text Detector module, which converts it to textual information, if it is found in the image, and passes it to the ingredient list filtering module (see the fragment of the code in Fig. 6).

The text from the React Native Text Detector is passed to the user along with the filtering result: (1) a positive response that the product / product belongs to a diet if no prohibited ingredients were found, or (2) unwanted ingredients are highlighted in the text, with an explanatory diet a list of the diets for which the prohibited ingredients were found in the list of ingredients of the product. When such ingredients are found, they are highlighted in red in the result view displayed in the list of ingredients, accompanied by a list of diets for which prohibited ingredients were found in the product. The data flow diagram of the ingredient filtering module is provided in Fig. 5.

The users' activity in the app is shown in Fig. 7, where only the views that can be accessed by an authorized user are covered.

The application database runs in the cloud and takes up less than 80Mb on the user's device, while user-captured images are cached on the app without saving them to the smart device's file system.

```javascript
import React, {useState, useEffect, useContext} from 'react';
import {Text, View, TouchableOpacity, ActivityIndicator, Image, Alert} from 'react-native';
import {RNCamera} from 'react-native-camera';
import RNTextDetector from 'react-native-text-detector';
import styles from './styles';
import firestore from '@react-native-firebase/firestore';
import '../../assets/i18n';
import {useTranslation} from 'react-i18next';
import {AuthContext} from '../../../navigation/AuthProvider';
const CameraScreen = ({route, navigation}) => {
  const {user, setUser} = useContext(AuthContext);
  const [loading, getLoading] = useState(false);
  const [photoUri, getPhotoUri] = useState([]);
  const [customIngredients, getCustomIngredients] = useState([]);
  const [userDiets, getUserDiets] = useState([]);
  let camera;
  const {t} = useTranslation();

  useEffect(() => {
    if (user) {
      const subscriber = firestore()
        .collection('Users')
        .doc(user.uid)
        .onSnapshot((doc) => {
          if (doc.data()) {
            getUserDiets(doc.data().chosen_diets || []);
            getCustomIngredients(doc.data().custom_unwanted_ingredients || []);
          }
        });
      return subscriber;
    } else return [];
  }, []);
  const takePicture = async (camera) => {
    if (camera) {
      const options = {quality: 0.8, base64: true};
      await camera.takePictureAsync(options).then((data) => {
        getPhotoUri(data.uri);
        getLoading(true);
        processDocument(data.uri);
      });
    }
  };
  const tryAgain = () => {
    Alert.alert(t('noTextAlert'));
    getPhotoUri(null);
    getLoading(false);
  };
  //converts the text found in the image to a text string
  const processDocument = async (photoUri) => {
    const textArray = await RNTextDetector.detectFromUri(
      photoUri,
    ).then((data) => data.map((textSnippet) => textSnippet.text));
    //converts the received array of text fragments into a single text string (in lowercase letters and separated by commas)
    if (textArray && Array.isArray(textArray) && textArray.length > 0) {
      const text = textArray.reduce((accumulator, text) => {
        return accumulator + text + ',';
      });
      const ingredients = text.toLowerCase().split(',');
      getForbiddenIngredients(ingredients, photoUri);
    } else {
      tryAgain();
    }
  };

  //Returns the prohibited ingredients of the user-selected diet in an object with the name of the diet and its prohibited ingredients
  const getForbiddenIngredients = async (ingredients, photoUri) => {
    let forbiddenIngredientObjects = [];
    if (userDiets.length === 0) {
      forbiddenIngredientObjects.push({
        diet: 'Custom',
        forbiddenIngredients: customIngredients,
      });
    } else {
      const a = await firestore()
        .collection('Diets')
        .where('name', 'in', userDiets)
        .get()
        .then((doc) =>
          doc.docs.map((doc) => {
            forbiddenIngredientObjects.push({
              diet: t(doc._data.name),
              forbiddenIngredients: doc._data.forbidden_ingredients,
            });
          }),
        )
        .then(() =>
          forbiddenIngredientObjects.push({
            diet: 'Custom',
            forbiddenIngredients: customIngredients,
          }),
        );
    }
    filterIngredients(ingredients, forbiddenIngredientObjects, photoUri);
  };

  //finds prohibited ingredients in the ingredient list of the product & finds diets that have the ingredient
  const filterIngredients = (
    ingredients,
    forbiddenIngredientObjects,
    photoUri,
  ) => {
    let ingredientsToBeRendered = [];
    let forbidden = '';
    let foundDiets = [];
    ingredients.map((ingredient, ingredientIndex) => {
      forbiddenIngredientObjects.map((object) => {
        object.forbiddenIngredients.map((forbiddenIngredient) => {
          //If a forbidden ingredient is found in an ingredient, writes it in the forbidden variable & the diet - in the diet list (if not already there)
          if (ingredient.includes(forbiddenIngredient)) {
            forbidden = forbiddenIngredient;
            if (!foundDiets.includes(object.diet)) foundDiets.push(object.diet);
            return;
          }
        });
        if (forbidden != '') return;
      });
```

**Fig. 6.** Code fragment of image processing and determination of forbidden ingredients

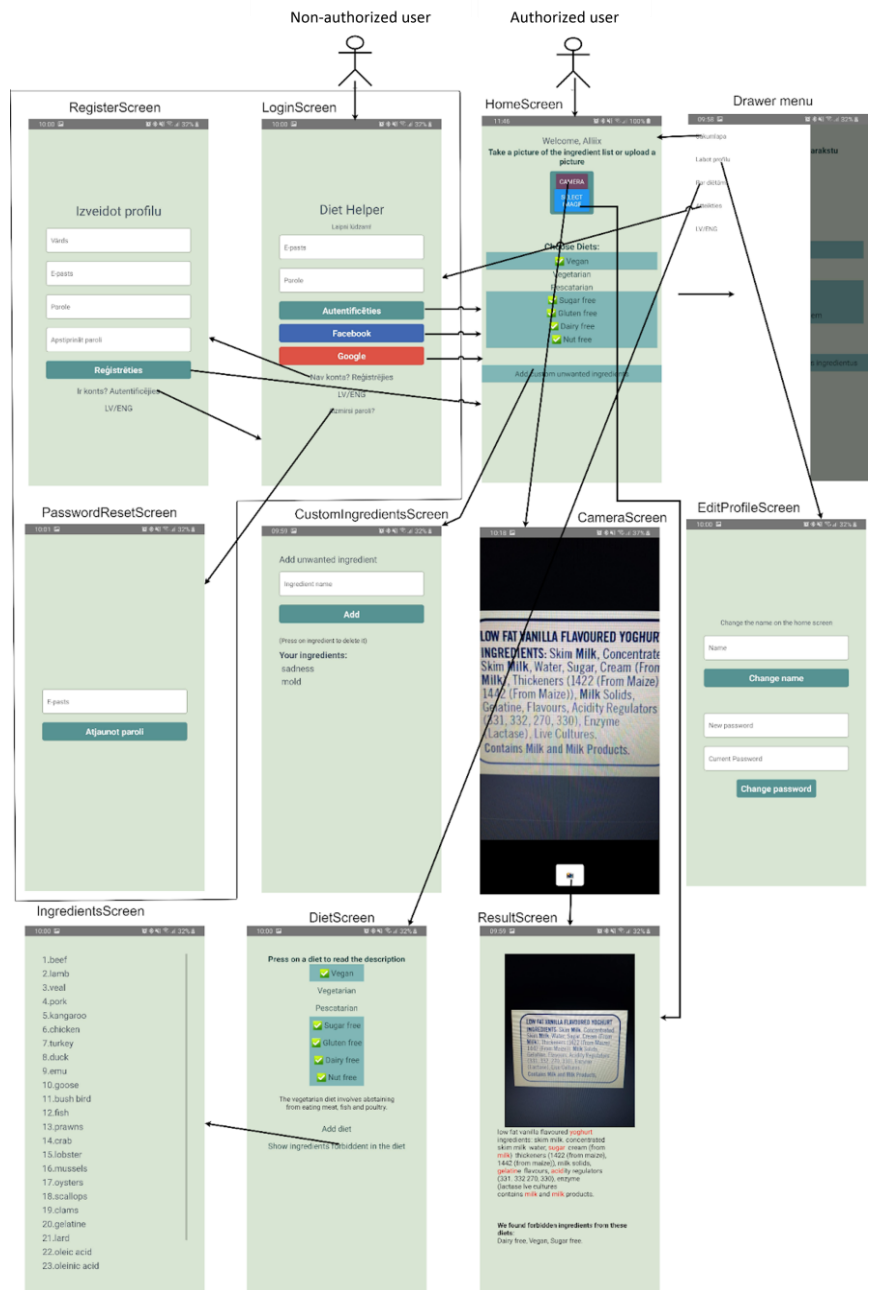

**Fig. 7.** Users' activity flow

The developed app was tested for compatibility with various *Android* smart devices with different API levels using the *Firebase Device Test Lab*, where a total of 7 manufacturers' smart devices, 8 *Android* versions (Kitkat 4.4.x, Jelly Bean 4.3.x, Lollipop 5.0.x, Lollipop 5.1.x, Marshmallow 6.0.x, Nougat 7.0.x, Nougat 7.1.x, Oreo 8.0.x, Oreo 8.1.x, Pie 9.x, Q 10.x), and 12 API levels were tested, as well as conducting acceptance testing, testing of both modules and user interface testing, in both cases passing all defined tests successfully.

## Limitations and Future Works

There are several limitations forming the future work. Firstly, the developed application is designed for *Android OS* smart devices. However, it can be extended to work on *iOS* smart devices since the code written in *React Native* can be relatively easily adapted for use in *Apple* products.

Secondly, currently, the app's ingredient filtering system works in English, thus covering the widest range of products, where ingredient lists are more common in English, but in the future the app's ingredient analysis feature may be available in several languages. This can be achieved by using the

React Native Text Detector, which recognizes characters in the Latin alphabet that are also used in other languages. In this case, the database can be supplemented with translations of the ingredients into the added languages.

Third, the scope of this app was food ingredients, however it is clear that the database of the application could be expanded to cover not only food ingredients but also, for example, household or cosmetic products, which may also be of undesirable origin or cause an allergic reaction.

Fourth, it is expected to ensure the user a possibility to use the app without an internet connection in the future. This is expected to be done by saving the lists of ingredients in a local database to search for them without using the web, thereby also saving the smartphone's battery. However, at the same time this would require storing the whole database locally, which should rather become an option, which the user may (de)select, not the only option.

And fifth, the database of diets and respective ingredients included in them is currently static, where the data are retrieved from the above-mentioned sources combined together. In the future, if the respective data could be found in an open data format, it could be made more dynamic. This would allow to ensure fast updates in the app if the diet is updated in one of the sources used, i.e., new item added or a particular ingredient is proved to be not the allergen, although the list of ingredients included in the diets are not a subject for often changes.

And the last but not the least is a potential of this app to be used to track the safety of the product, indicating the list of ingredients or foods hazardous to humans, thereby contributing to the food safety being a part of Sustainable Development Goals (SDG). This is also the movement gaining popularity in the world, particularly in the context of food safety (e.g., The European Food Safety Authority (EFSA) activity), but not only. This would require database extension with the list of ingredients found to be hazardous to humans – here a constantly updated database mentioned above is more crucial prerequisite.

## Conclusions

The paper presents a simple technological solution to a complex problem that affects both the social, emotional, and financial wellbeing of a large part of society – a mobile shopping assistant that processes ingredient label transforming it into the text (by means of *React Native*), analyse the list of food ingredients and filter unwanted ingredients that the user should avoid due to either allergies or personal reasons. The system is based on a meaningful and sufficiently complete database that combines data on different diets and their respective ingredients from three different sources, thus forming an enriched database. However, the limited availability of open access resources, i.e., open data being publicly accessible and well-maintained, prevents from dynamic updating of the database. Although the data required by *Diet Helper* do not change frequently, this can become an issue, when its scope extended, including food safety assessment, which requires a constantly updated database.

The app allows users to personalize the experience of the application by identifying and defining unwanted / prohibited ingredients by themselves. In this way *Diet Helper* digital shopping assistant represents m-health domain and is expected to reduce the time spent by people/ consumers in groceries with more accurate and simplified product selection and filtering in seconds, thereby improving the social, emotional, and financial wellbeing of those affected by restrictions on the consumed products or their ingredients.

The proposed system is simple enough to allow similar solutions to be developed for other domains with a high level of risk to humans' health, social and emotional wellbeing etc. even with limited resources. In other words, the paper is an encouragement and even call to start actively developing solutions that could improve our lives.